\documentclass[twocolumn,preprintnumbers,amsmath,amssymb]{revtex4}
\usepackage{amsmath,amssymb,graphics,epsfig,subfigure}
\usepackage{color}

\begin{document}

\thispagestyle{empty}

\begin{center}

\title{Thermodynamic curvature of the Schwarzschild-AdS black hole and Bose condensation}

\date{\today}
\author{Sandip Mahish\footnote{E-mail: sm19@iitbbs.ac.in}, Aritra Ghosh\footnote{E-mail: ag34@iitbbs.ac.in} and Chandrasekhar Bhamidipati\footnote{E-mail: chandrasekhar@iitbbs.ac.in}}

\affiliation{School of Basic Sciences, Indian Institute of Technology Bhubaneswar,\\   Jatni, Khurda, Odisha, 752050, India}

\begin{abstract}
In the AdS/CFT correspondence, a dynamical cosmological constant $\Lambda$ in the bulk corresponds to varying the number of colors $N$ in the boundary gauge theory with a chemical potential $\mu$ as its thermodynamic conjugate. In this work, within the context of Schwarzschild black holes in $AdS_5 \times S^5$ and its dual finite temperature $\mathcal{N}=4$ superconformal Yang-Mills theory at large $N$, we investigate thermodynamic geometry through the behavior of the Ruppeiner scalar $R$. The sign of $R$ is an empirical indicator of the nature of microscopic interactions and is found to be negative for the large black hole branch implying that its thermodynamic characteristics bear qualitative similarities with that of an attraction dominated system, such as an ideal gas of bosons. We find that as the system's fugacity approaches unity, $R$ takes increasingly negative values signifying long range correlations and strong quantum fluctuations signaling the onset of Bose condensation. On the other hand, $R$ for the small black hole branch is negative at low temperatures and positive at high temperatures with a second order critical point which roughly separates the two regions.

\end{abstract}
\maketitle
\end{center}

\section{Introduction}
Although thermodynamics is built on a macroscopic framework without {\em per se} any reference to the microscopic physics, it is robust enough to yield deep insights into a wide range of phenomena which principally originate at the microscopic level. Thus, it has remained as an indispensable tool in our long standing efforts to learn about the quantum nature of black holes and gravity, where a full understanding of the microscopic degrees of freedom is still lacking. Black hole thermodynamics commenced with the seminal papers of Bekenstein and Hawking~\cite{Bekenstein:1972tm, Bekenstein:1973ur, Hawking:1974sw} on relating the entropy $S$ to the area $A$ and the temperature \(T\) to the surface gravity \(\kappa\) of the event horizon. This was then followed by the formulation of the laws of black hole mechanics in analogy with those of thermodynamics \cite{Bardeen:1973gs} paving the way for a thorough investigation of the quantum structure of black holes. The presence of a cosmological constant $\Lambda$ makes the thermodynamic structure of black holes richer with the emergence of phase transitions~\cite{Hawking:1982dh,Chamblin}. Today, investigations on both microscopic and macroscopic aspects of black holes, particularly in theories with a negative $\Lambda$ are quite fascinating and often lead to a mutual understanding of the underlying physics of quantum gravity as well as that of quantum field theories through the gauge/gravity duality \cite{Maldacena:1997re,Witten:1998zw}.

\smallskip

An interesting aspect of black hole thermodynamics ensues while generalizing Komar's definition of mass from asymptotically flat to (A)dS spacetimes, namely the requirement of a dynamical cosmological constant. A variable $\Lambda$ plays a central role in extending Smarr's formula from flat to AdS spacetimes, typically giving rise to the notion of pressure $P$ and a novel concept of thermodynamic volume $V$~(see \cite{Kastor:2009wy,Cvetic:2010jb} and references therein for some earlier attempts). This novel set up where the first law is augmented by a $P dV$ term coming from a dynamical $\Lambda$ is known as extended black hole thermodynamics. In this approach, the mass $M$ of the black hole is identified with its enthalpy (rather than internal energy $U$) as $H=M=U + PV$ in the bulk with the new first law being written as,
\begin{equation} \label{FLbulk}
dH = T dS + V dP \, ,
\end{equation}
where $V= \frac{\partial H}{\partial P}\Bigr|_S$ is the thermodynamic volume~\cite{Cvetic:2010jb} conjugate to $P$. An important advantage of the availability of $P$ and $V$ terms is the possibility of writing an equation of state for black holes in analogy with the familiar hydrostatic systems~\cite{Dolan:2010ha}. For charged black holes, there turns out to be an exact identification of its phase transitions~\cite{Hawking:1982dh,Chamblin} (including the matching of critical exponents) with the well known liquid-gas transitions of a van der Waals fluid (vdW)~\cite{Kubiznak:2012wp} putting them in the same universality class.

\smallskip

A negative $\Lambda$ in the bulk is generally associated with the degrees of freedom of the dual conformal field theory (CFT) on the boundary. Since a dynamical cosmological constant leads to the introduction of a pressure in bulk, it is interesting to ask whether the same interpretation is applicable to the boundary CFT as well. Examining this interpretation in the case of $AdS_5 \times S^5$ and the finite temperature ${\mathcal N}=4$ superconformal Yang-Mills theory at large \(N\) reveals that it is more suitable to view a dynamical $\Lambda$ as being associated to varying number of colors~\cite{Kastor:2009wy,Dolan:2013dga,Johnson:2014yja,Caceres:2015vsa,K,dolan} of the dual CFT. To see this, let us consider the set up used in~\cite{dolan} when there is a black hole in the bulk. In $AdS_5 \times S^5$, the line element corresponding to a Schwarzschild black hole is~\cite{Chamblin},
\begin{equation}
\label{10d}
ds^2=ds_{AdS_5}^2 + l^2\,d\Omega_{5}^2,
\end{equation}
where $d\Omega_{5}^2$ is the line element on a five dimensional sphere with unit radius. Here, $ds_{AdS_5}^2$ is the line element of the Schwarzschild-AdS black hole given as,
\begin{equation} \label{bh}
 ds_{AdS_5}^2 = -f(r) \, dt^2 + f(r)^{-1} \, dr^2 + r^2\, d\Omega_3^2,
\end{equation} with \(d\Omega_3^2\) being the metric on a 3-sphere. The lapse function \(f(r)\) reads,
\begin{equation}
  f(r) = 1 - \frac{8G_{(5)}M}{3\pi r^2} + \frac{r^2}{l^2},
\end{equation}
where \(l\) is the radius of the \(AdS_5\) spacetime related to the cosmological constant as $\Lambda = -\frac{6}{l^2}$. Here, $M$ is the black hole mass and the five dimensional Newton's constant \(G_{(5)}\) appearing in the black hole solution is not fixed but is tied to $l$ as,
\begin{equation} \label{bulk}
\frac{1}{16\pi G_{(5)} } = \frac{\pi^2 l^2}{16 G_{(10)}}\, .
\end{equation}
Here, it is the ten dimensional Newton's constant $G_{(10)}$ and the ten dimensional Planck length $l_P$ (linked as $\hbar G_{(10)} = l^8_P$) which are held fixed. The ten-dimensional spacetime [eqn (\ref{10d})] can be obtained from $N$ coincident $D3$-branes in type IIB supergravity in the near horizon limit. In that situation, the AdS radius $l$ is related to the number of D3-branes $N$ as~\cite{Maldacena:1997re},
\begin{equation}\label{boundary}
l^4 = \frac{\sqrt{2} N}{\pi^2}l^4_P \, .
\end{equation}
As per the AdS/CFT correspondence, the spacetime [eqn (\ref{10d})] is thought of as the gravity dual of $\mathcal{N} = 4$ superconformal Yang-Mills (SYM) theory at finite temperature and large \(N\) with $N$ being the rank of the gauge group of the $SU(N)$ SYM theory. Although $\Lambda$ is taken to be fixed in arriving at above relations, in the full ten dimensional bulk solution, $\Lambda$ is not a priori known and is just a parameter like any other constant in the supergravity solution which can be dynamically generated through the vacuum expectation values of the scalar fields~\cite{Freedman:1999gp,Myers:2012ed} and therefore can be varied. Based on the relations given in eqns (\ref{bulk}) and (\ref{boundary}), it is clear that as long as $N$ is large \footnote{$N \rightarrow \infty$ in the classical limit and can be taken to be a variable, like the particles in a gas are treated in the thermodynamic limit.}, it can vary if $\Lambda$ (and hence $l$) varies since $l_P$ is held fixed. A varying $N$ can either come from exchanging the color degrees of freedom with a bath or can instead be viewed as a dialing constant while still corresponding to a dynamical $\Lambda$ in the bulk. The thermodynamic conjugate to $\Lambda$ on the boundary should then correspond to a chemical potential $\mu$ for the number of colors $N$~\cite{dolan}, with the first law in this context written as (keeping in mind that in the large $N$ limit, the number of degrees of freedom on the boundary scales as $N^2$: coming from massless bosons and fermions in the weak coupling limit~\cite{Gubser:1998nz}),
\begin{equation}
dM = T dS + \mu dN^2 \, .
\end{equation}
The chemical potential $\mu$ can be calculated explicitly as,
\begin{equation} \label{mu1}
\mu = \frac{\partial M}{\partial N^2}\Bigr|_S \, .
\end{equation}
In statistical physics, chemical potential contains important information about the system, as it is negative for an ideal Bose gas and positive only if there is a sufficiently strong repulsive interaction among the particles or in Fermi systems at low temperatures where the exclusion principle plays a central role~\cite{callen}. In particular, the chemical potential can be zero when the average thermal de Broglie wavelength of particles is comparable to the inter-particle separation signifying the onset of strong quantum effects in the system besides signifying loss of particle number conservation. Remarkably, in the context of Schwarzschild black holes in $AdS_5 \times S^5$ and ${\mathcal N}=4$ SYM at large $N$, $\mu$ was computed in~\cite{dolan} (also see \cite{cai1,cai2,wei,Belhaj:2015uwa,Kastor:2014dra,Kastor:2016bph}) from eqn (\ref{mu1}) and it was noted that there is a point where it goes to zero too. The temperature at which it approaches zero is close to the point where the black hole undergoes the Hawking-Page (HP) transition and it was proposed to correspond to Bose condensation. This was improved further in~\cite{sarkar}, where it was shown that $\mu=0$ might be reached in other systems involving charged, rotating and higher curvature AdS black holes, exactly at or above the HP transition temperature if one makes use of densities instead of absolute thermodynamic quantities.

\smallskip

To ascertain the existence of Bose condensation~\cite{dolan}, it is important to investigate whether there are regions in the thermodynamic phase space where the system shows characteristics reminiscent of Bose statistics. The aim of this work is to study the behavior of the thermodynamic curvature or Ruppeiner scalar \footnote{Throughout this paper, we shall interchangeably use the terms thermodynamic curvature and Ruppeiner scalar.} by using thermodynamic quantities in terms of densities rather than absolute quantities and to develop a novel thermodynamic understanding of the system by isolating the contributions of the large and small black hole branches. From the analysis of the Ruppeiner scalar in terms of temperature $T$ and fugacity $z$, we find that the large black hole branch has a remarkable resemblance with an ideal Bose gas with the possibility of Bose condensation as $z \rightarrow 1$. On the other hand, we infer empirically from the sign of the Ruppeiner scalar that the small black hole branch is attraction dominated at low temperatures and repulsion dominated at high temperatures with a second order critical point approximately separating the two regions.

\section{Thermodynamics with chemical potential} \label{chem}
The black hole mass can be obtained to be \cite{dolan},
\begin{equation}\label{M}
  M = \frac{3\pi r_+^2 }{8G_{(5)}l^2}\bigg(1 + \frac{r_+^2}{l^2}\bigg).
\end{equation} On the boundary, the mass is set equal to the internal energy, i.e. \(U = M\) \cite{Johnson:2014yja}. The Bekenstein-Hawking entropy is,
\begin{equation}\label{S}
  S = \frac{\pi^2 r_+^3}{2G_{(5)}} = \frac{\pi^5 l^5 r_+^3}{2 l_P^8}.
\end{equation}
To set up thermodynamics with chemical potential in the context of AdS/CFT correspondence, it is useful to use $S$ and $N^2$ as thermodynamic variables, in place of the quantities intrinsic to black holes, such as $r_+$ and $l$. Further, following \cite{cai2,sarkar} we consider the densities of the extensive thermodynamic quantities \(U\) and \(S\) by scaling them down by a factor of \(l^3\) (as the volume of the dual CFT goes as \(\sim l^3\)). The energy density \(u\) is then given as,
\begin{equation} \label{rhodensity}
  u = \frac{3 s^{2/3} \left(N^{5/6}+2 \sqrt[3]{2} s^{2/3}\right)}{4\ 2^{2/3} \pi  N^{2/3}}.
\end{equation}
It satisfies the first law of black hole thermodynamics expressed in terms of densities as,
\begin{equation}\label{fl}
  du = Tds + \mu dN^2,
\end{equation} where \(s\) is the entropy density obtained from eqn (\ref{S}). The Hawking temperature is given by,
\begin{equation} \label{Tdensity}
T= \bigg(\frac{\partial u}{\partial s}\bigg)_{N^2} =\frac{N^{5/6}+4 \sqrt[3]{2} s^{2/3}}{2 \times 2^{2/3} \pi  N^{2/3} \sqrt[3]{s}}.
\end{equation}
It has a minimum value, \(T_{min}\) which occurs at \(s_0 =\frac{N^{5/4}}{8 \sqrt{2}}\). For any temperature above  \(T_{min}\), there are two values of \(s\) with the same temperature: \(s < s_0\) corresponds to the small black hole branch whereas \(s > s_0\) corresponds to the large black hole branch.

\subsection{Chemical potential}
\begin{figure}[h]
\begin{center}
\includegraphics[width=3.4in]{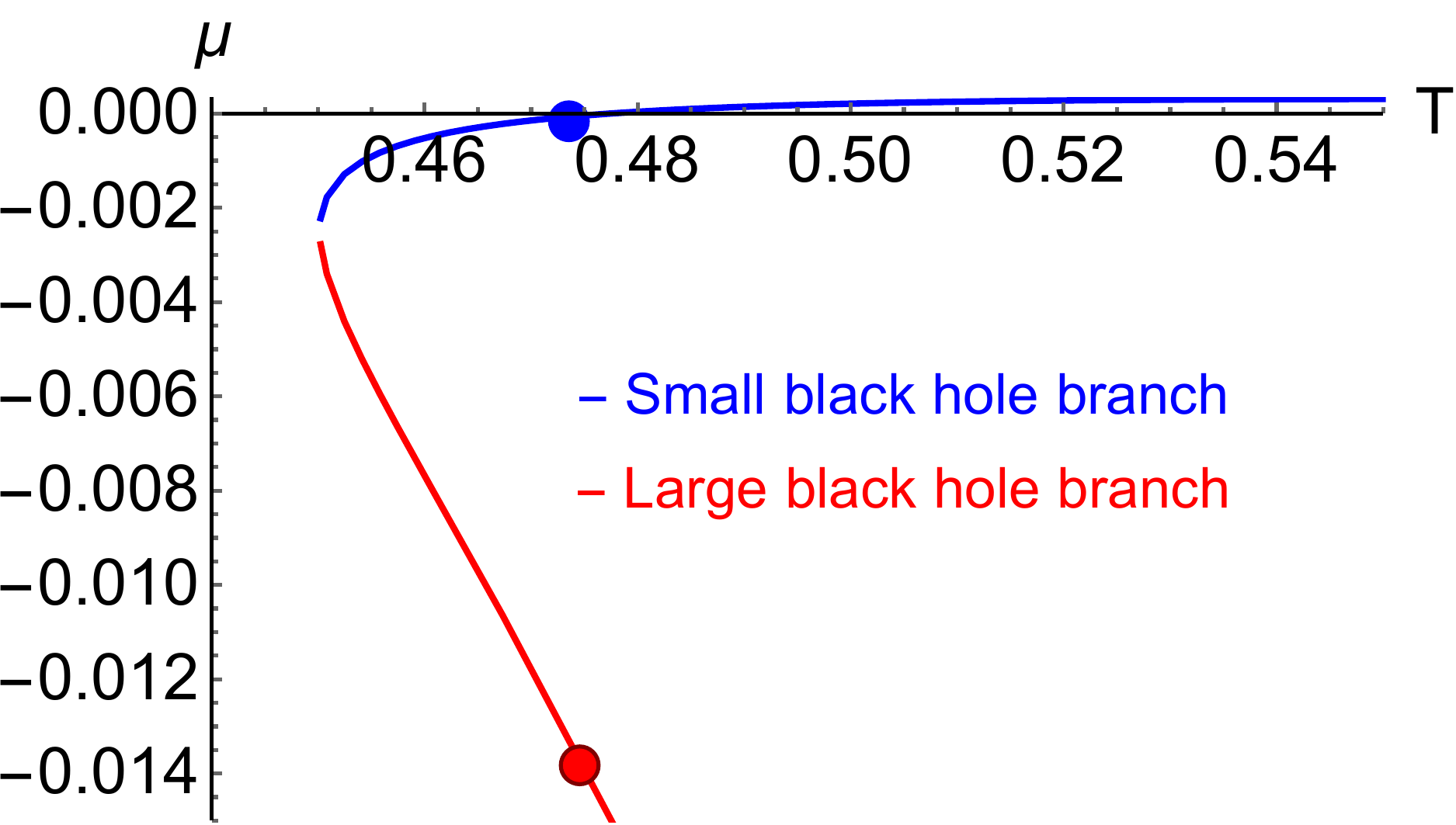}
\caption{Chemical potential of the AdS Schwarzschild black hole as a function of the Hawking temperature with fixed \(N^2\): HP transition (red dot in large black hole branch) and zero of chemical potential (blue dot in small black hole branch) happen at the same temperature $T=0.477465\, N^{-1/4}$.}\label{muplotT}
\end{center}
\end{figure}

Chemical potential can now be defined from the first law [eqn (\ref{fl})] as,
\begin{equation} \label{mudensity}
  \mu =: \bigg(\frac{\partial u}{\partial N^2}\bigg)_s = \frac{\sqrt[3]{2} N^{5/6} s^{2/3}-8 \times 2^{2/3} s^{4/3}}{32 \pi  N^{8/3}}\, ,
\end{equation}
which is plotted in figure-(\ref{muplotT}).
The chemical potential is negative definite in the regime corresponding to the large black hole branch and monotonically goes to negative infinity as \(T \rightarrow \infty\) (or equivalently \(s \rightarrow \infty\)). It can be checked that the chemical potential goes to zero exactly at the Hawking-Page temperature \(T_{HP} = \frac{3}{2 \pi  \sqrt[4]{N}}=0.477465 N^{-1/4}\) as also noted in \cite{sarkar}. The Hawking-Page transition occurs in the large black hole branch at entropy density \(s_{HP} = \frac{N^{5/4}}{4}\) and \(\mu\) goes to zero in the small black hole branch with \(s_{\mu=0} =\frac{N^{5/4}}{32}\), all at the same temperature. It is helpful to define fugacity as,
\begin{equation} \label{fugacity}
  z = e^{\frac{\mu}{T}},
\end{equation}
which gives a better picture during approach to quantum limit. Normally, for a quantum gas, \(z \rightarrow 0\) is the classical limit, whereas \(z \rightarrow 1\) is the limit where quantum effects become significant and for an ideal gas of bosons is also the indication of Bose condensation. With this in mind, let us analyze the behavior of the system in the regions corresponding to the small and large black hole branches in order.\\

\noindent

{\underline {Small black hole branch}}: There is a zero crossing at \(s_{\mu=0} = \frac{N^{5/4}}{32}\) such that for black holes smaller than \(s< s_{\mu=0}\), \(\mu\)  is positive and vice versa. In terms of the Hawking temperature, black holes colder than the \(\mu = 0\) case are associated with negative chemical potential whereas the hotter ones have positive chemical potentials. For \(T > T_{\mu=0}\), the fugacity is positive and close to unity while if \(T_{min} < T < T_{\mu=0}\), fugacity can get negative but is still close to unity.

\smallskip

{\underline {Large black hole branch}}:  The chemical potential being negative definite, i.e. \(-\infty > \mu > 0\)  is a characteristic of an ideal Bose gas, which will be argued in next section using Ruppeiner geometry to be coming from having interactions which are purely of attractive type, as noted in a different context~\cite{Janyszek1990}. The classical limit is reached by the large black hole as \(T \rightarrow \infty\) or equivalently \(\mu \rightarrow -\infty\). It should be noted that \(z \in (0,1)\).

\subsection{Behavior of the specific heats}
From the first law [eqn (\ref{fl})], one can obtain the specific heat of the system \cite{cai1} in terms of densities at constant \(N^2\) as,
\begin{equation}\label{CN2}
  C_{N^2} = \bigg(\frac{\partial u}{\partial T}\bigg)_{N^2} = -\frac{3 s \left(N^{5/6}+4 \sqrt[3]{2} s^{2/3}\right)}{N^{5/6}-4 \sqrt[3]{2}
   s^{2/3}} \, .
\end{equation}
It has a divergence at $T = T_{min}$ and is positive in the region of the large black hole while it is negative for that of the small black hole~\cite{Hawking:1982dh}. One can think of this feature as being reminiscent of a second order point much like that pointed out by Davies \cite{Davies} which separates the positive and negative regions of the specific heat in rotating black holes. In a grand canonical sense, of interest is the specific heat at constant chemical potential \(C_\mu\). It is calculated to be,
\begin{equation}\label{cmu}
C_{\mu} = \bigg(\frac{\partial \phi}{\partial T}\bigg)_{\mu} = -\frac{-\frac{512\ 2^{2/3} s^{7/3}}{N^{5/6}}+11 N^{5/6} s-84 \sqrt[3]{2}
   s^{5/3}}{3 N^{5/6}-36 \sqrt[3]{2} s^{2/3}},
\end{equation} where \(\phi = u - \mu N^2\). It has a divergence at $T_{C_\mu = \infty}=\frac{2 \sqrt{\frac{2}{3}}}{\pi  \sqrt[4]{N}}$ in the small black hole branch and can be understood as an isolated critical point of the small black hole.

\section{Thermodynamic geometry}
In equilibrium thermodynamics, Ruppeiner geometry is set up by taking the Boltzmann entropy $S$ as the key thermodynamic potential~\cite{Ruppeiner} with its Hessian (computed w.r.t. to other fluctuating thermodynamic variables) giving rise to the notion of a metric on the spaces of thermodynamic equilibrium states. The existence of such a metric endows spaces of thermodynamic equilibrium states with the notion of a length between the states such that, shorter the distance between a pair of thermodynamic states, the more probable is a fluctuation between them. The curvature scalar \(R\) associated with such a geometry can be calculated for a given thermodynamic system and turns out to contain information about phase transitions and critical points which has now been tested for different physical systems such as ideal gases, vdW fluids, 1-D Ising models, quantum Bose/Fermi gases etc \cite{Janyszek1990, Ruppeiner2, Janyszek:1989zz, Ruppeiner3}. The understanding which has developed from studying these systems is as follows. For the classical ideal gas \(R\) is zero, whereas it is negative definite for the van der Waals fluid suggesting a somewhat mysterious connection with the presence of intermolecular interactions. In the case of ideal quantum gases, the ideal Bose and Fermi systems turn out to have $R<0$ and $R>0$ respectively and approach $R \rightarrow 0$ as one takes the high temperature limit which is consistent with classical behavior. In a sense therefore, the sign of \(R\) acts as an empirical indicator of the nature of microscopic interactions such that \(R < 0\) suggests the dominance of attractive interactions whereas, \(R > 0\) suggests the dominance of repulsive interactions between the microscopic degrees of freedom. Further, it has been argued that the divergence of $R$ indicates a strongly correlated or a critical behavior of the microscopic degrees of freedom (see for example \cite{RuppeinerRMP}), whereas the absolute value of $R$ gives a measure of the strength of interactions~\cite{Ruppeinerinteractions}. Therefore, Ruppeiner geometry is a macroscopic probe which can be used to find out (at least empirically) the nature of interactions of a thermodynamic system, unlike more conventional approaches in statistical physics where the starting point is a microscopic model. For black hole systems where the underlying microscopic picture is unclear, Ruppeiner geometry can be expected to provide physical insights about the overall dominant interactions of the microscopic degrees of freedom. To the best of our knowledge, Ruppeiner geometry in black hole thermodynamics was first studied in \cite{Cai} and more recently it is being applied to several black hole systems \cite{Wei2015,Dehyadegari, AR, Wei:2019uqg, Wei2019b, Xu, GBRuppeiner, GBWei, BTZ, Dark, Naveen, Xu2}.

\smallskip

\begin{figure}[h]
\begin{center}
\includegraphics[width=3.4in]{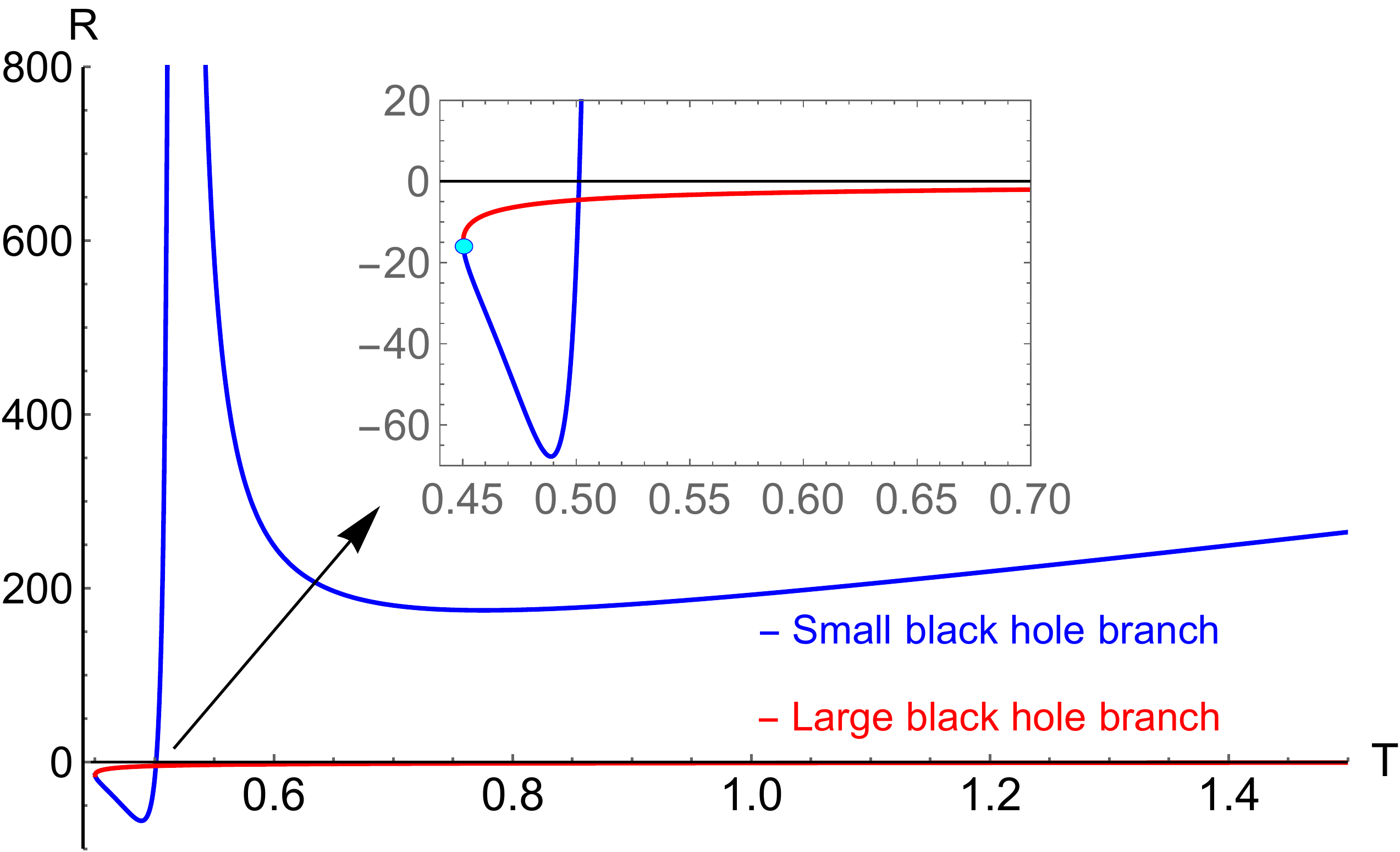}
\caption{Ruppeiner scalar $R$ of the AdS Schwarzschild black hole as a function of temperature $T$ with fixed \(N^2\). {\em Inset:} blow up of the region in the plot showing (a) the point (blue dot) where the large and small black hole branches germinate in AdS, (b) that $R$ is always negative for the large black hole branch (curve in red) and (c) the point where $R$ for the small black hole branch (blue curve) has a zero crossing.}\label{RplotT}
\end{center}
\end{figure}

We note that the Ruppeiner metric is defined as the negative Hessian of the entropy (here entropy density) so that the corresponding line element is given by \cite{Ruppeiner,Ruppeiner2,RuppeinerRMP,Ruppeinerinteractions},
\begin{equation}\label{me}
  dl_R^2 = - \frac{\partial^2 s}{\partial x^i \partial x^j} dx^i dx^j,
\end{equation}where \(i,j \in \{1,2,....,n\}\) and \(\{x^i\}\) are independent coordinates allowed to fluctuate on the space thermodynamic equilibrium states. Using the first law given in eqn (\ref{fl}), the metric can be expressed as~\cite{cai1,cai2,wei} (see Appendix for exact derivation),
\begin{equation}\label{RuppeinerSN}
  dl_R^2 = \frac{1}{C_{N^2}}(ds)^2 + \frac{2}{T}\bigg(\frac{\partial T}{\partial N^2}\bigg)_s(ds)(dN^2) +\frac{1}{T}\bigg(\frac{\partial \mu}{\partial N^2}\bigg)_s(dN^2)^2.
\end{equation}
In this case, the fluctuation coordinates are taken to be \(s\) and \(N^2\). The metric is two dimensional, i.e. with the generic structure \(dl_R^2 = g_{11} (dx^1)^2 + g_{12} (dx^1)(dx^2) + g_{21} (dx^2)(dx^1) + g_{22} (dx^2)^2\) with \(g_{12}= g_{21}\). In this situation, the curvature scalar corresponding to the geometry described by the metric can be obtained to be \cite{R},
\begin{eqnarray}
    R = - \frac{1}{\sqrt{g}}\bigg[\frac{\partial}{\partial x^1}\bigg(\frac{g_{12}}{g_{11}\sqrt{g}}\frac{\partial g_{11}}{\partial x^2} - \frac{1}{\sqrt{g}}\frac{\partial g_{22}}{\partial x^1}\bigg)  \nonumber \\
 + \frac{\partial}{\partial x^2}\bigg(\frac{2}{\sqrt{g}}\frac{\partial g_{12}}{\partial x^2} - \frac{1}{\sqrt{g}}\frac{\partial g_{11}}{\partial x^2} - \frac{g_{12}}{g_{11}\sqrt{g}}\frac{\partial g_{11}}{\partial x^1}\bigg) \bigg],
\end{eqnarray}
where \(g=g_{11}g_{22}-g_{12}g_{21}\) is the determinant of the metric tensor. Using this, the thermodynamic curvature or Ruppeiner scalar in terms of \(s\) and \(N^2\) as the fluctuating variables can be straightforwardly calculated and reads,
\begin{equation}\label{Rsn}
R = \frac{A}{B},
\end{equation} where,
\begin{eqnarray}
  A = 8 (40\ 2^{2/3} N^{5/3} s^{2/3}+160 N^{5/6} s^{4/3} \nonumber \\
     -5 \sqrt[3]{2} N^{5/2}+768 \sqrt[3]{2} s^2),\\
  B = 3 N^{5/6} \sqrt[3]{s} \left(N^{5/6}-12 \sqrt[3]{2} s^{2/3}\right)^2 \nonumber \\
   \times \left(N^{5/6}+4 \sqrt[3]{2} s^{2/3}\right).
\end{eqnarray}
The Ruppeiner scalar expressed in terms of absolute quantities does not have any zero crossings~\cite{cai1}, but in terms of densities, it can be shown from eqn (\ref{Rsn}) that there is a novel zero crossing, which is meaningful for the small black hole. This zero crossing is clearly visible after isolating the behavior of the large and small black hole branches, as shown in figure-(\ref{RplotT}). Below we discuss the contributions of both black hole branches in sequel.\\

\noindent

{\underline {Large black hole branch}}: For the large black hole, $R$ in eqn (\ref{Rsn}) is found to be always negative suggesting heuristically the possible dominance of attractive interactions in this system. In fact, at very high temperatures we have,
\begin{equation}\label{RT}
R_{T\rightarrow \infty} =- \frac{32}{9 \pi  N^{3/2} T}-\frac{8}{3 \pi ^3 N^2 T^3} + \cdots \, ,
\end{equation}
which approaches zero.
The vanishing of $R$ is a signature that the system is approaching a point where the interactions between the degrees of freedom are negligible. Furthermore, both $R$ and \(\mu\) are negative definite (also see figure-(\ref{muplotT})) in this branch. Hence, it is pragmatic to model the large black hole as an ideal gas with bosonic degrees of freedom. The high temperature values of thermodynamic quantities can be calculated to check this. For instance, energy and entropy densities respectively go as,
\begin{equation} \label{rhoT}
u_{T\rightarrow \infty} = \frac{3}{16} \pi ^3 N^2 T^4  -\frac{3}{16} \pi  N^{3/2} T^2-\frac{3 \sqrt{N}}{64 \pi ^3 T^2}-\frac{3 N}{32 \pi }+ \cdots \, ,
\end{equation}
\begin{equation} \label{sT}
s_{T\rightarrow \infty}=\frac{1}{4} \pi ^3 N^2 T^3 -\frac{3}{8} \pi  N^{3/2} T-\frac{\sqrt{N}}{32 \pi ^3 T^3}+ \cdots \, .
\end{equation}
The system at very high temperatures in the CFT is deep in the de-confined phase. In~\cite{Gubser:1996de}, thermodynamic quantities of non-extremal black 3-branes in type IIB supergravity were shown to match (up to certain factors) with independent statistical mechanical computations performed in the dual CFT, assuming a dilute gas of $3+1$ dimensional massless open string states. The leading terms in the energy and entropy density in the high temperature limit shown in eqns (\ref{rhoT}) and (\ref{sT}) respectively, are same (up to numerical factors) as in~\cite{Gubser:1996de,Aharony:1999ti} and satisfy the relation $s \sim N^{1/2}\, u^{3/4}$~\cite{Aharony:1999ti}. It is interesting that the Ruppeiner scalar (\ref{RT}) also approaches zero, making a correct prediction of a non-interacting system in the high temperature limit where the system can be modeled as a gas of $N^2$ free particles in $(3+1)$ dimensions~\cite{Gubser:1996de}.
\begin{figure}[h]
\begin{center}
\includegraphics[width=3.1in]{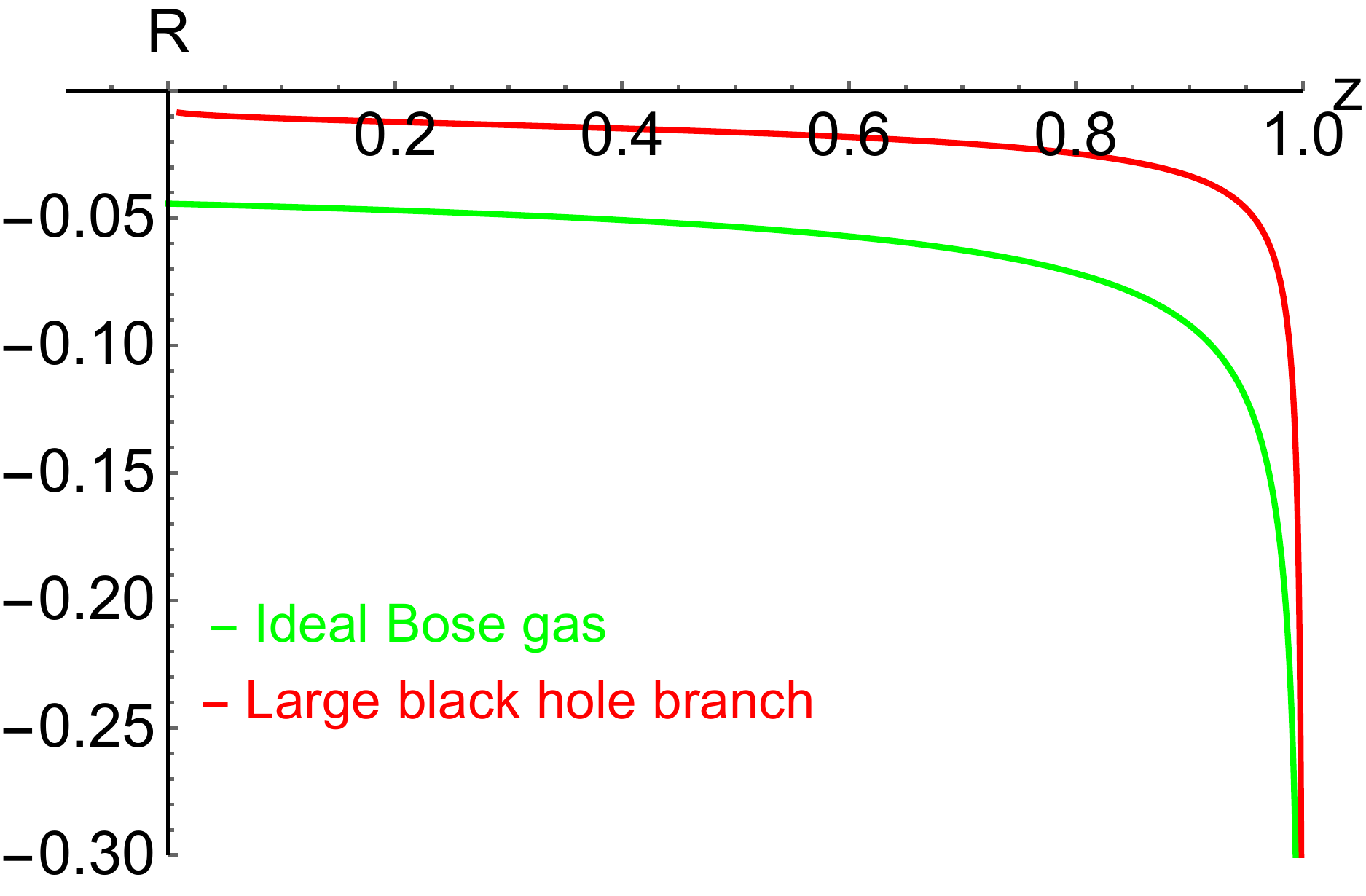}
\caption{Comparative plots of Ruppeiner scalar $R$ as a function of fugacity $z$ for the large black hole branch (in red, scaled by 0.01) and ideal Bose gas~\cite{Janyszek1990} (in green). For both the cases, fugacity has the same range, i.e., from $0<z<1$ and $R$ is always negative and continues to decrease monotonically as $z \rightarrow 1$.}\label{Rz}
\end{center}
\end{figure}
Having empirically associated attractive interactions and ideal Bose behavior to the region corresponding to the large black hole branch, one can now check for Bose condensation. To investigate this issue, it is useful to express the Ruppeiner scalar $R$ in terms of fugacity introduced in eqn (\ref{fugacity}) and study its features, particularly, close to the $z=1$ region. As the expressions are quite complex, the result is plotted in figure-(\ref{Rz}) (red curve). In the high temperature limit, earlier observations of last paragraph hold that $R \rightarrow 0$ as fugacity \(z \rightarrow 0\); At low temperatures, as \(z\) approaches unity, \(R\) monotonically decreases and rapidly takes large negative values. Ruppeiner scalar obtained in~\cite{Janyszek1990} in terms fugacity \footnote{Ruppeiner scalar for an ideal Bose gas is~\cite{Janyszek1990},
$$
R = - \frac{g^2_{3/2}(z)g_{1/2}(z)-2g_{5/2}(z)g^2_{1/2}(z)+g_{5/2}(z)g_{3/2}(z)g_{-1/2}(z)}{\left( 5g_{5/2}(z)g_{1/2}(z) - 3 g^2_{3/2}(z)\right)^2} \, ,
$$
where we have ignored overall factors and inserted a negative sign to match our sign conventions. $g_{\nu}(z)$ stands for Bose-Einstein functions.} is plotted in figure-\ref{Rz}) (green curve) in comparison with the large black hole branch. The behavior of $R$ for the large black hole branch is seen to be remarkably similar to the one computed long back by Janyszek and Mrugala in~\cite{Janyszek1990} for an ideal Bose gas, using a microscopic statistical mechanical partition function. Basing on the nearly identical behavior of both systems over the entire range of variation of fugacity, as seen from figure-\ref{Rz}, the limit \(z \rightarrow 1\) can thus be taken to indicate the onset of Bose condensation, supporting the proposal put forward in~\cite{dolan}.\\

\noindent

{\underline {Small black hole branch}}:
Coming to the small black hole branch, there are a few distinct features of the Ruppeiner scalar which are interesting to note from figure-(\ref{RplotT}). First, \(R\) starts out as being positive at high temperatures and then exactly at a critical temperature \(T_c = 0.519798  N^{-1/4}\), it diverges, which is also the point where the specific heat \(C_\mu\) diverges. This point can be interpreted as a second order critical point around which both the specific heat \(C_\mu\) and the Ruppeiner scalar \(R\) can be shown to have scaling behavior. Particularly, upon defining \(t = T/T_c-1\), the specific heat goes at \(t \rightarrow 0\) as,
\begin{equation}
  C_\mu \sim |t|^{-1},
\end{equation} which gives the value of the critical exponent \(\alpha = 1\). It is also straightforward to show that the Ruppeiner scalar exhibits the following scaling as \(t \rightarrow 0\),
\begin{equation}
  R \sim |t|^{-2},
\end{equation}
which matches with the ones noted earlier~\cite{wei,Wei:2019uqg,Kumar:2014uba,Maity:2015rfa} in slightly different contexts. Consequently, we expect that the correlation volume also scales with temperature around the critical point \cite{RuppeinerRMP}. Next, further at a slightly lower temperature \(T_{R = 0}=0.501256 N^{-1/4}\), the Ruppeiner scalar crosses zero. The two above temperatures differ approximately by \(3.56 \%\) only. $R$ is thus negative for the temperature range \(T_{R = 0} > T > T_{min}\). Therefore, from an analysis of the Ruppeiner scalar, it might be reasonable to presume that the behavior of the small black hole branch is akin to a system with repulsive interactions at high temperatures and attractive interactions at low temperatures with a switching between the two regions taking place at the temperature \(T = T_{R=0}\) and the second order critical point at \(T=T_c\) approximately separating the two regions. While all of this happens in the \(\mu > 0\) region, one may attempt to qualitatively explain this behavior by postulating the small black hole branch to be associated with both attractive and repulsive interactions consisting of gas with dual statistics~\cite{mirzaAnyon} (see also \cite{AR,GBRuppeiner}). One may further speculate that at a given temperature, the relative number densities of microstructures of the two kinds may dictate which kind of interaction would be dominant in the system~\cite{Wei2015}. It would be interesting to explore these suppositions.

\section{Discussion}
In this work, we studied the thermodynamic geometry of the Schwarzschild-AdS black hole in $AdS_5 \times S^5$ in the context of the AdS/CFT correspondence where a dynamical cosmological constant in the bulk corresponds to varying number of colors in the dual $\mathcal{N}=4$ superconformal Yang-Mills theory at finite temperature and large $N$. We learnt from the analysis of the thermodynamic curvature or Ruppeiner scalar $R$ that the supposed behavior of the large black hole branch greatly mimics that of an ideal Bose gas also showing the characteristics of Bose condensation (figure-\ref{Rz}). Also, just like the large black hole behavior of the Ruppeiner scalar in the high temperature limit was argued from eqn (\ref{RT}) through a CFT input~\cite{Gubser:1996de}, an understanding of the small black hole branch would be interesting to pursue, possibly following new developments~\cite{Hanada:2016pwv}. Furthermore, an independent computation of thermodynamic curvature on boundary CFT has not been attempted before and would be interesting to pursue. It should be borne in mind that Ruppeiner geometry is entirely based on a macroscopic formulation and therefore an exact picture of the deduced nature of microstructure interactions and Bose condensation in this system is a subject of further investigation which has to be supplemented by an independent microscopic analysis. Here, it is pertinent to mention that Ruppeiner geometry and the conclusions drawn from the nature (and sign) of $R$ are not a substitute for computations done through microscopic counting of states in string theory and other models of gravity. Still, knowing the nature of $R$ is helpful in gaining early insights into the nature of dominant interactions (at least empirically) at a given temperature or a given horizon radius or in a particular parameter regime for black holes.

\section*{Acknowledgements}
One of us (C.B.) would like to thank Sudipta Mukherji for earlier discussions and collaboration on topics related to the subject of this work, during visits to Institute of Physics, Bhubaneswar. We thank B. P. Dolan and S. -W. Wei for their valuable comments on the manuscript. Finally, we thank the referee for valuable comments which improved the manuscript.

\section*{Appendix: Derivation of the Ruppeiner line element on the \((s,N^2)\)-plane}
In order to derive the line element of the Ruppeiner metric [eqn (\ref{me})] with independent coordinates \(s\) and \(N^2\), let us start out with the generic expression \(dl_R^2 = -g_{\mu \nu} dx^\mu dx^\nu\) where \(g_{\mu \nu} = \partial_\mu \partial_\nu s \). Writing out \(dz_\mu = g_{\mu \nu} dx^\nu\) one gets,
\begin{equation}\label{1}
  dl_R^2 = - dz_\mu dx^\mu.
\end{equation}
Since,
\(dz_\mu = g_{\mu \nu} dx^\nu\), we must have,
\begin{equation}
  z_\mu = \frac{\partial s}{\partial x^\mu}.
\end{equation}
Now, from the first law given in eqn (\ref{fl}), one can write,
\begin{equation}\label{firstlawcontact}
  ds - \frac{du}{T} + \frac{\mu dN^2}{T} = 0
\end{equation}
which means that \(z_1 = 1/T\) and \(z_2 = - \mu/T\) whereas \(x^1 = u\) and \(x^2 = N^2\) such that \(ds = z_\mu dx^\mu\). Then, with these identifications,
\begin{equation}
  dz_1 = -\frac{dT}{T^2}, \hspace{2mm} dz_2 = \frac{\mu dT}{T^2} - \frac{d\mu}{T}.
\end{equation} The line element given in eqn (\ref{1}) which has the form \(dl_R^2 = - dz_2 dx^2 - dz_2 dx^2\) can now be written as,
\begin{equation}
  dl_R^2 = - \bigg(-\frac{dT}{T^2}\bigg) du - \bigg(\frac{\mu}{T^2}dT - \frac{d\mu}{T}\bigg) dN^2,
\end{equation} which from the first law reduces to,
\begin{equation}\label{lineelementgeneric1}
  dl_R^2 =  \frac{dsdT}{T} + \frac{d \mu dN^2}{T}.
\end{equation} At this stage, if one considers \(s\) and \(N^2\) to be independent coordinates such that,
\begin{equation}
  T = T(s,N^2), \hspace{3mm} \mu = \mu(s,N^2),
\end{equation} then differentiating these and substituting them into eqn (\ref{lineelementgeneric1}) one finally gets,
\begin{equation}\label{Ruppeinerlineelementfinal}
  dl_R^2 = \frac{1}{C_{N^2}}(ds)^2 + \frac{2}{T}\bigg(\frac{\partial T}{\partial N^2}\bigg)_s(ds)(dN^2) +\frac{1}{T}\bigg(\frac{\partial \mu}{\partial N^2}\bigg)_s(dN^2)^2,
\end{equation}
where we have used the (Maxwell-like) relation,
\begin{equation}
  \bigg(\frac{\partial T}{\partial N^2}\bigg)_s = \bigg(\frac{\partial \mu}{\partial s}\bigg)_{N^2},
\end{equation} thus completing the derivation. It can be checked that the line element so obtained is nothing but the Weinhold metric \cite{Weinhold} defined as a Hessian of the energy (in this case, energy density \(u\)) together with an overall conformal factor of inverse temperature, i.e. \(T^{-1}\).

\end{document}